\documentclass[aps,pra,showpacs,twocolumn,superscriptaddress]{revtex4}
\usepackage{amsmath}
\usepackage{amsfonts}
\usepackage{amssymb}
\usepackage{graphicx}
\usepackage{bbm}
\usepackage{natbib}
\usepackage[dvipdfm]{hyperref}
\makeatletter

\begin{document}

\title{Quantum Zeno and anti-Zeno effects in quantum dissipative systems}

\author{Wei Wu}

\email{weiwu@csrc.ac.cn}

\affiliation{Beijing Computational Science Research Center, Beijing 100193, People's Republic of China}

\author{Hai-Qing Lin}

\affiliation{Beijing Computational Science Research Center, Beijing 100193, People's Republic of China}

\begin{abstract}
We investigate the quantum Zeno and anti-Zeno effects in quantum dissipative systems by employing a hierarchical equations of motion approach which is beyond the usual Markovian approximation, the rotating wave approximation, and the perturbative approximation. The quantum Zeno and anti-Zeno dynamics of a biased qubit-boson model and a biased qutrit-boson model are provided as illustrative examples. It is found that (i) there exists multiple Zeno-anti-Zeno crossover phenomena, (ii) the non-Markovian characteristic of the bath may be favorable for the accessibility of the Zeno dynamics, and (iii) high bath temperature may add the difficulty in observing the quantum Zeno effect in quantum dissipative systems.
\end{abstract}
\pacs{03.65.Xp, 03.67.Yz, 42.50.Dv}
\maketitle

\section{Introduction}\label{sec:sec1}

The quantum Zeno effect (QZE) describes a quantum phenomenon that the decay of an unstable quantum system can be frozen or suitably confined by repeated frequent measurements~\cite{1}. On the other hand, the reverse effect, i.e., the acceleration of the decay process of the unstable quantum system induced by repeated frequent measurements, has also been pointed out and is known as the quantum anti-Zeno effect (QAZE)~\cite{2}. The QZE has been experimentally observed in many real physical systems, such as the trapped ion~\cite{3}, the superconducting Josephson junction~\cite{4}, the ultracold atomic Bose-Einstein condensate~\cite{5}, and the nuclear spin system~\cite{6}.

On the other hand, due to the unavoidable coupling with the surrounding bath, the microscopic quantum system severely undergoes decoherence
which is the main difficulty in fulfilling reliable quantum computation and quantum communication tasks~\cite{7,8}. In this sense, to gain a global view and more physical insights into the QZE and the QAZE, the effect of the surrounding bath should be taken into consideration. Almost all the existing studies of the QZE and the QAZE in quantum dissipative systems have restricted their attentions to some exactly solvable models, such as the pure dephasing model~\cite{9,10} and the unbiased qubit-boson model with rotating-wave approximation~\cite{10,11}, which is also called the damped Jaynes-Cummings model. For the unbiased qubit-boson model beyond the rotating-wave approximation~\cite{12}, most of the treatments are based on the generalized Silbey-Harris transformation~\cite{12,13} which is valid only in the weak system-bath coupling regime. Very few studies focus on the more general quantum dissipative systems, such as the biased qubit-boson model~\cite{14,15,16} where both the dephasing mechanism and quantum relaxation are considered in this model. Many previous articles have shown that the existence of an external bias field can remarkably change the decoherence behavior of the qubit-boson model~\cite{14,15}. Thus we expect that the QZE and the QAZE of a biased qubit-boson model is quite different from that of the unbiased case. Unfortunately, the reduced dynamics of the biased qubit-boson model cannot be exactly obtained, in this paper, we employ a numerical hierarchical equations of motion (HEOM) method~\cite{17,18,19,20} to study the QZE and the QAZE in quantum dissipative systems.

The HEOM is a set of time-local differential equations for the reduced density matrix of the quantum subsystem, which was originally proposed by Tanimura and his co-workers~\cite{17,18}. This numerical treatment includes all the orders of the system-bath interactions and is beyond the usual Markovian approximation, the rotating wave approximation, and the perturbative approximation. In recent years, the HEOM approach was successfully used to study the reduced dynamics in many chemical and biophysical systems, such as optical line shapes of molecular aggregates and electron energy transfer dynamics in the Fenna-Matthews-Olson complex~\cite{17}. Furthermore, the HEOM method is also employed to investigate some important problems in the field of the quantum information science, such as the dynamics of entanglement or quantum discord in quantum dissipative systems~\cite{20}, as well as the dynamical behaviors of the spin squeezing and the quantum Fisher information under certain non-Markovian decoherence channels~\cite{19,21}.

During the past years, there has been an increasing interest to study the memory effect of the bath in quantum dissipative systems~\cite{22,23,24,25}; this memory effect, which is also known as the non-Markovianity, is a very important characteristic of the quantum dissipative dynamics and has many applications in realistic physical systems~\cite{25}. Since the non-Markovian bath retains the memory and has some feedback action on the quantum subsystem, the reduced dynamics of the quantum subsystem could be considerably changed. It would be of great interest in an investigation of the QZE and the QAZE in a non-Markovian bath. Moreover, we are not only interested in the QZE or the QAZE in the non-Markovian bath, but also in the modifications of the QZE (or the QAZE) induced by the non-Markovianity. What is the link between the non-Markovianity and the QZE (the QAZE)? Or more specifically, what is the influence of the non-Markovianity on the QZE or the QAZE in quantum dissipative systems? In this paper, we try to address this question by studying the QZE and the QAZE of a biased qubit (and a qutrit) coupled to a zero temperature non-Markovian bosonic bath.

On the other hand, for practical quantum devices, the influences of bath temperature on the decoherence can not be disregarded. One commonly believed concept is that the bath temperature can speed up the destruction of quantum coherence. However, some studies have shown that the bath temperature is able to reduce the decoherence in some quantum dissipative systems~\cite{26,27}. This result is contrary to the common recognition that a higher bath temperature always induces a more severe decoherence and suggests that the bath temperature plays a very intricate role in quantum dissipative systems. An interesting question arises here: what is the influences of the bath temperature on the QZE or the QAZE in quantum dissipative systems? To address the above concern, we also generalize our study to the finite temperature situation.

This paper is organized as follows. In Sec.~\ref{sec:sec2}, we briefly outline some basic concepts as well as the general formalism of the QZE and the QAZE in a general quantum dissipative system. In Sec.~\ref{sec:sec3}, we study the QZE and the QAZE in some general spin-boson models at both zero and finite temperatures by the numerical HEOM approach. Some concerned discussions and the main conclusions of this paper are drawn in Sec.~\ref{sec:sec4}.

\section{The QZE and QAZE in a general quantum dissipative system}\label{sec:sec2}

In this section, we would like to make a brief summary of the main features of the QZE and the QAZE in quantum dissipative systems. The general Hamiltonian of a quantum dissipative system can be described by
\begin{equation}\label{Eq:Eq1}
\hat{H}=\hat{H}_{s}+\hat{H}_{b}+\hat{H}_{sb},
\end{equation}
where $\hat{H}_{s}$ is the Hamiltonian of the quantum subsystem, $\hat{H}_{b}$ denotes the Hamiltonian of the surrounding bath and $\hat{H}_{sb}$ stands for the system-bath interaction Hamiltonian. In our work, we regard the quantum subsystem $\hat{H}_{s}$ as the object to be measured. Throughout, we work in the dimensionless units and $\hbar=k_{B}=1$. Assuming the initial state of the quantum subsystem is $\hat{\varrho}_{s}(0)=|\psi(0)\rangle\langle\psi(0)|$ and the bath is prepared in a thermal equilibrium state $\hat{\varrho}_{b}=\exp(-\beta\hat{H}_{b})/\mathrm{Tr}_{b}[\exp(-\beta\hat{H}_{b})]$, where $\beta\equiv T^{-1}$ is the inverse temperature, then the reduced density matrix of the quantum subsystem at time $t$ is given by $\hat{\varrho}_{s}(t)=\mathrm{Tr}_{b}[\exp(-i\hat{H}t)\hat{\varrho}_{s}(0)\otimes\hat{\varrho}_{b}\exp(i\hat{H}t)]$.

Suppose that the quantum subsystem is probed $N$ times by the projective measurements $\mathcal{\hat{M}}=|\psi(0)\rangle\langle\psi(0)|$ with equal time intervals $\tau=t/N$ during its time evolution, the survival probability after the
measurements is given by
\begin{equation}\label{Eq:Eq2}
\begin{split}
P(t)=&P(\tau)^{N}\\
=&[\langle\psi(0)|\exp(i\hat{H}_{s}\tau)\hat{\varrho}_{s}(\tau)\exp(-i\hat{H}_{s}\tau)|\psi(0)\rangle]^{N},
\end{split}
\end{equation}
where we have applied a time-dependent rotation $\exp(i\hat{H}_{s}\tau)$ to remove the evolution induced by $\hat{H}_{s}$ itself, this treatment is extensively adopted in many previous studies~\cite{9,28} and would be very helpful for us to study the QZE and QAZE induced by the dissipative bath. In our theoretical formalism, we assume that the state of the bath is not disturbed by the measurements, namely, the bath is always in the thermal equilibrium state after each measurement on the quantum subsystem. We also would like to mention that it might be interesting to explore the quantum Zeno and anti-Zeno effects of a quantum subsystem embedded in a non-equilibrium environment. The study in this field is beyond the the scope of our paper and needs further investigation.

For the sake of convenience, one can introduce the effective decay rate which is defined by
\begin{equation}\label{Eq:Eq3}
\Gamma(\tau)\equiv-\frac{1}{\tau}\ln[P(\tau)].
\end{equation}
Then the survival probability after the measurements can be rewritten as $P(t)=\exp[-\Gamma(\tau)t]$. The effective decay rate $\Gamma(\tau)$ is the crucial physical quantity to study the QZE and the QAZE in quantum dissipative systems~\cite{1,2,9,10,12,14,28}. In many previous studies, the ratio of $\Gamma(\tau)/\Gamma_{0}$ is used to identify the occurrence of the QZE and the QAZE, where $\Gamma_{0}$ is the natural decay rate of the quantum subsystem obtained by the Fermi golden rule~\cite{1,2,12}. In this paper, we adopt an alternative way to characterize the QZE and the QAZE~\cite{9,10,28,29}: the QZE takes place when the effective decay rate $\Gamma(\tau)$ decreases as $\tau$ becomes smaller, while, the QAZE occurs when the effective decay rate $\Gamma(\tau)$ increases as $\tau$ becomes smaller. The local maximum or minimum of $\Gamma(\tau)$ is the transition point between the QZE regime and the QAZE regime. This definition of QZE and QAZE has a clear physical picture: if the rapidly repeated measurements decrease the value of effective decay rate, the relaxation process of the measured system is suppressed which leads to the QZE, on the other hand, if the rapidly repeated measurements increase the value of effective decay rate, the relaxation process of the measured system is accelerated which leads to the QAZE. When $\Gamma(\tau)=0$, neither the QZE nor the QAZE occurs. The classification of the QZE and the QAZE by the behaviors of $\Gamma(\tau)$ is very suitable for the case where the natural decay rate is unknown~\cite{9,10,28,29}.

The other important physical quantity to describe the QZE is the quantum Zeno time. One of the most widely used definitionS of the quantum Zeno time is given by~\cite{1,12}
\begin{equation}\label{Eq:Eq4}
\tau_{\mathrm{Z}}\equiv \Bigg{[}\frac{d}{d\tau}\Gamma(\tau)\Bigg{]}^{-\frac{1}{2}}_{\tau\rightarrow 0}.
\end{equation}
With the help of this definition, the survival probability can be approximately expressed as $P(t)=1-t^{2}/\tau_{\mathrm{Z}}^{2}$ in the short-time regime. One can find that a larger value of $\tau_{\mathrm{Z}}$ makes the QZE easier to realize~\cite{10,30}.

\section{Results}\label{sec:sec3}

The dissipation-induced decoherence in a quantum microscopic system can be effectively modeled by the spin-boson model, which describes the interaction between a spin subsystem and a bosonic bath. The spin-boson model has attracted considerable attention in past decades because it provides a very simple model to simulate many physical and biological processes. The reduced system dynamics of the spin-boson model has been studied by various analytical and numerical methods, for example, the generalized Silbey-Harris transformation approach~\cite{12,13,14}, the time-dependent density matrix renormalization group method~\cite{31}, the Dirac-Frenkel time-dependent variation with Davydov ansatz~\cite{32}, and the HEOM formalism~\cite{17,18,19,20}.

In this paper, we consider a general spin-boson model which is described by the following Hamiltonian:
\begin{equation}\label{Eq:Eq8}
\hat{H}=\hat{H}_{s}+\sum_{k}\omega_{k}\hat{a}_{k}^{\dagger}\hat{a}_{k}+f(\hat{s})\otimes \sum_{k}g_{k}(a_{k}^{\dagger}+a_{k}),
\end{equation}
where $f(\hat{s})$ denotes the quantum subsystem's operator coupled to the bath, $\omega_{k}$ is the frequency of the $k$th boson mode, $g_{k}$ labels the coupling strength between the spin and the $k$th boson mode, and $\hat{a}_{k}$ and $\hat{a}^{\dag}_{k}$ are the annihilation and creation operators of the $k$th boson mode, respectively. We use the HEOM method to investigate the QZE and the QAZE in this general spin-boson model. The explicit expressions of the hierarchical equations at both zero and finite temperatures are given in the Appendix. In this section, the QZE and QAZE of a biased qubit-boson model and a biased qutrit-boson model are provided as illustrative examples. In order to demonstrate the accuracy of our numerical results, we also compare our numerical results with the exactly analytical results in the pure dephasing cases and the perturbative results proposed in Ref.~\cite{28}.

\subsection{The biased qubit-boson model}\label{sec:sec3a}

\begin{figure}
\centering
\includegraphics[angle=0,width=6cm]{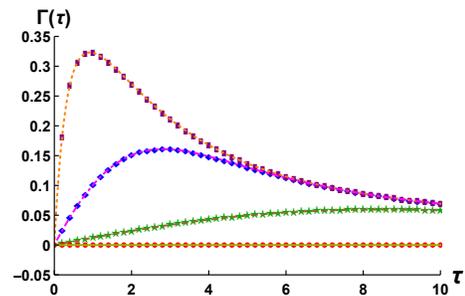}
\caption{\label{fig:fig1} Effective decay rate $\Gamma(\tau)$ of the pure dephasing qubit-boson model obtained by the numerical HEOM method and the exact analytical expression for initial state $|\psi(0)\rangle=\frac{1}{\sqrt{2}}(|e\rangle+|g\rangle)$ with different parameters: $\lambda=10\gamma_{0}$ (numerical result: orange dotted line, analytical result: purple rectangles), $\lambda=\gamma_{0}$ (numerical result: magenta dotdashed line, analytical result: blue diamonds) and $\lambda=0.1\gamma_{0}$ (numerical result: brown dashed line, analytical result: green five-point stars). The yellow solid line and the red circles are the numerical and exact analytical results of initial state $|\psi(0)\rangle=|e\rangle$, respectively. Other parameters are chosen as $\epsilon=1$, $\gamma_{0}=0.5$, $\Delta=0$, and $\omega_{0}=0$.}
\end{figure}

In this subsection, we study the QZE and the QAZE in a biased qubit-boson model, the Hamiltonian of which is described by Eq.~\ref{Eq:Eq8} with $\hat{H}_{s}=\frac{\epsilon}{2}\hat{\sigma}_{z}-\frac{\Delta}{2}\hat{\sigma}_{x}$ and $f(\hat{s})=\hat{\sigma}_{z}$. The QZE and the QAZE in an unbiased qubit-boson model ($\epsilon=0$) have been studied in many previous articles~\cite{12}. There are several reasons why we need to consider the effect of a finite external bias field in our study. First, the existence of the finite external bias field enriches the type of the decoherence mechanism, both quantum dephasing ($\Delta=0$ and $\epsilon\neq 0$) and quantum relaxation ($[\hat{H}_{s},f(\hat{s})]\neq0$) cases are included in the same model. Second, some studies showed that the finite external bias field is able to enhance the quantum coherence~\cite{14,15}, this effect of the nonzero external bias field maybe helpful to realize the QZE or the QAZE in experiments, because the observation of the QZE or the QAZE requires a relatively long quantum relaxation or decoherence time~\cite{9}.

In this subsection, we assume the bath is initially prepared in its Fock-vacuum state $\bigotimes_{k}|0_{k}\rangle$ and the bath density spectral function $J(\omega)\equiv\sum_{k}g_{k}^{2}\delta(\omega-\omega_{k})$ has the Lorentz spectrum form
\begin{equation}\label{Eq:Eq9}
J_{\mathrm{L}}(\omega)=\frac{1}{2\pi}\frac{\gamma_{0}\lambda^{2}}{(\omega-\omega_{0})^{2}+\lambda^{2}},
\end{equation}
where $\lambda$ defines the spectral width of the coupling and $\gamma_{0}$ can be approximately interpreted as the system-bath coupling strength. The reason why we choose the Lorentz spectrum at zero temperature case is twofold: firstly, for a Lorentzian bath density spectral function $J_{\mathrm{L}}(\omega)$, $C_{\mathrm{L}}(t)$ (see Eq.~\ref{Eq:Eq17} in the Appendix) is the Ornstein-Uhlenbeck-type bath correlation function~\cite{7,33} which is the key requirement to perform the HEOM scheme~\cite{17,18,19,20}. Secondly, the Lorentzian spectrum has a clear boundary between Markovian and non-Markovian regimes~\cite{7,33}. More specifically speaking, the parameter $\lambda$ is connected to the bath correlation time $\tau_{b}$ by the relation $\tau_{b}\simeq \lambda^{-1}$, while the time scale $\tau_{s}$, on which the state of the system changes, is given by $\tau_{s}\simeq \gamma_{0}^{-1}$. In this sense, the boundary between Markovian regimes and non-Markovian regimes can be approximately specified by the ratio of $\tau_{b}/\tau_{s}=\gamma_{0}/\lambda$. When $\gamma_{0}/\lambda$ is very small, which means the bath correlation time $\tau_{b}$ is much smaller than the relaxation time of the quantum subsystem $\tau_{s}$, the decoherence mechanics is Markovian~\cite{7,33}. When $\gamma_{0}/\lambda$ is large, the memory effect of the bath should be taken into account and the dynamics of this open system is then non-Markovian~\cite{7,33}. In fact, one can demonstrate that the hierarchical equations in Eq.~\ref{Eq:Eq18} reduce to the usual Markovian Lindblad-type master equation in the limit $\lambda\gg\rm max \{\gamma_{0}, \omega_0\}$~\cite{18}, due to the fact that the bath correlation function reduces to the Dirac $\delta$ function $C_{\mathrm{L}}(t-t')\rightarrow \mathrm{const}.\times\delta(t-t')$ in this situation. This feature of the Lorentz spectrum is very helpful for us to study the relationship between the non-Markovianity and the QZE in quantum dissipative systems. However, we also want to point out that a more rigorous way to distinguish the Markovian or non-Markovian regimes in parameter space should consider not only the bath density spectral function $J(\omega)$ but also the decoherence channel. In recent years, many physical quantities, such as trace distance~\cite{22}, quantum Fisher information and quantum correlation~\cite{24}, are proposed to identify the non-Markovianity in quantum open systems. Unfortunately, only for a few models, one can obtain an analytical expression of the non-Markovianity. For the more general quantum open systems, these physical quantities or schemes are too hard to compute. It would be very interesting to establish a more rigorous relation between these non-Markovianity measures and the QZE in quantum dissipative systems.

For the pure dephasing case, namely $\Delta=0$, this model can be exactly solved. Assuming the initial state of the qubit is $|\psi(0)\rangle=\frac{1}{\sqrt{2}}(|e\rangle+|g\rangle)$, where $|e\rangle$ ($|g\rangle$) is the excited (ground) state of the Pauli $\hat{\sigma}_z$ matrix, one can obtain the explicit expression of the effective decay rate as follows~\cite{9}:
\begin{equation}\label{Eq:Eq10}
\Gamma(\tau)=-\frac{1}{\tau}\ln[\frac{1}{2}+\frac{1}{2}e^{-\kappa(\tau)}],
\end{equation}
where
\begin{equation*}
\kappa(\tau)=4\int d\omega J_{\mathrm{L}}(\omega)\frac{1-\cos(\omega\tau)}{\omega^{2}}.
\end{equation*}

In Fig.~\ref{fig:fig1}, we plot the effective decay rate $\Gamma(\tau)$ obtained by the numerical HEOM technique as well as the exactly analytical expression. A perfect agreement is found between the two different approaches. One can immediately observe that the effective decay rate $\Gamma(\tau)$ has a peak structure, which means there is a crossover between the QZE and the QAZE regimes. On the other hand, if the qubit is initially prepared in $|\psi(0)\rangle=|e\rangle$, one can find that the value of the effective decay rate $\Gamma(\tau)$ is zero which means neither the QZE nor the QAZE occurs in this case. This phenomenon can be easily understood because $|e\rangle$ is a dark state in this pure dephasing model and does not evolve. These results convince us that the numerical HEOM technique truly captures the quantum Zeno or the anti-Zeno dynamics in quantum dissipative systems.

\begin{figure}
\centering
\includegraphics[angle=0,width=4cm]{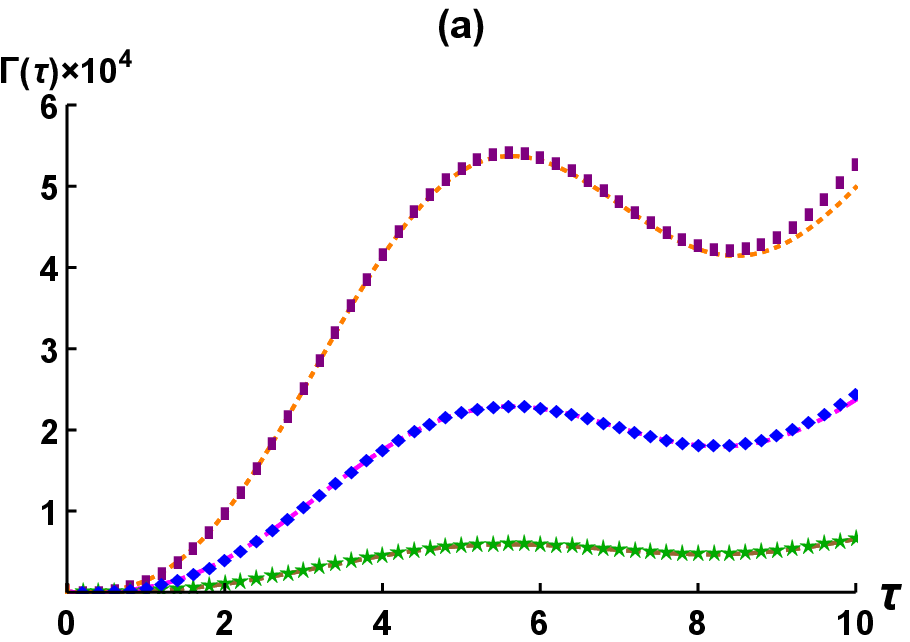}
\includegraphics[angle=0,width=4cm]{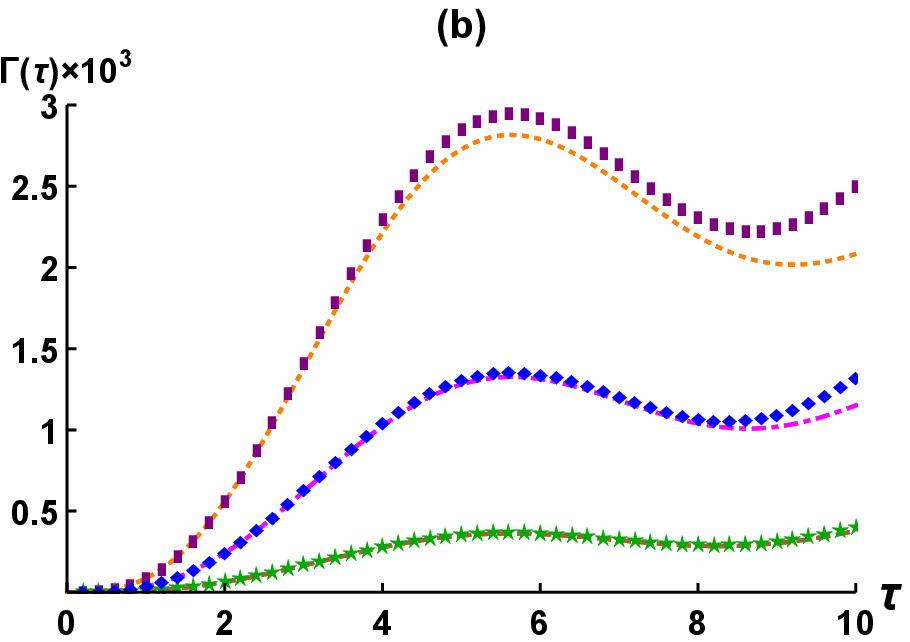}
\includegraphics[angle=0,width=4cm]{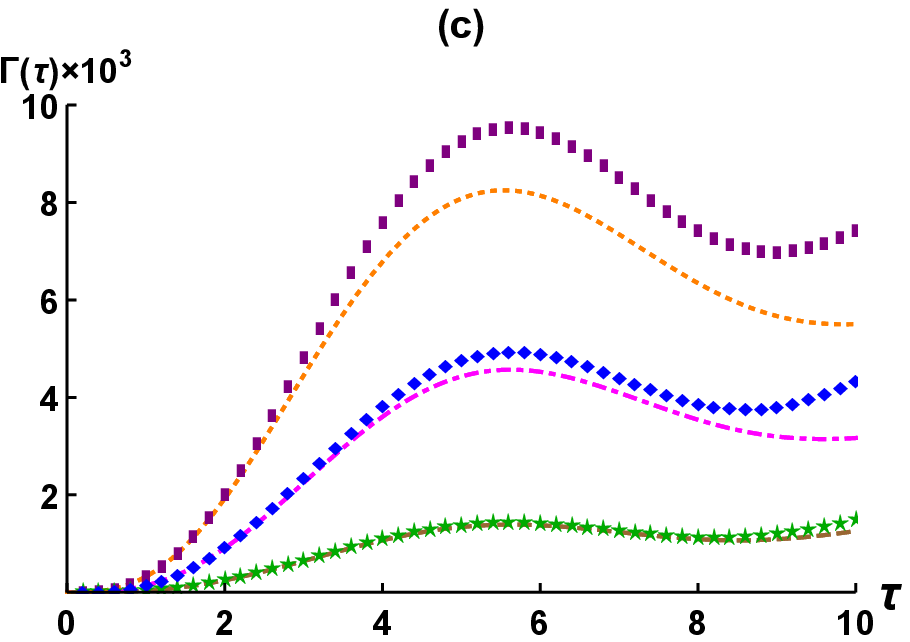}
\includegraphics[angle=0,width=4cm]{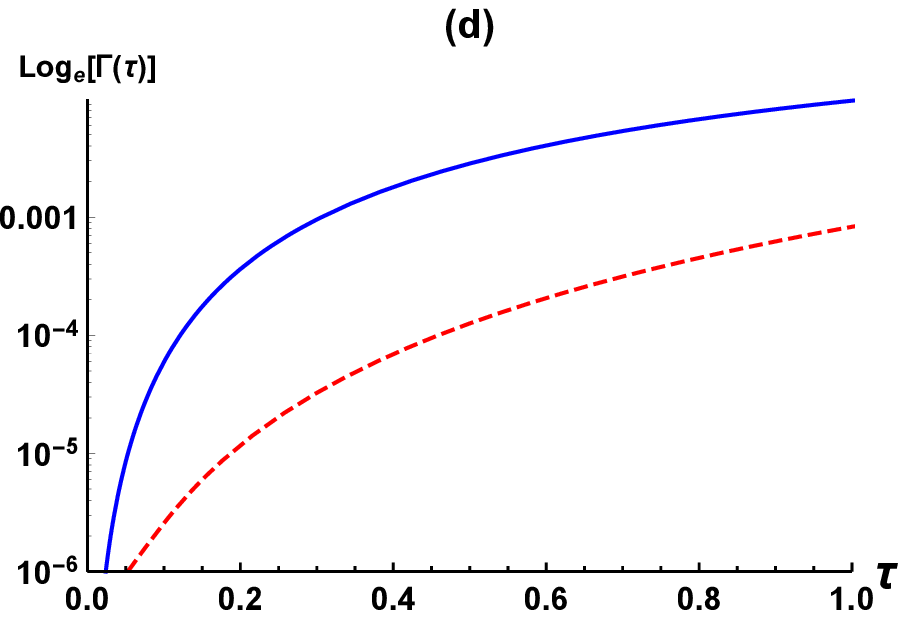}
\caption{\label{fig:fig2} (a) The effective decay rate $\Gamma(\tau)$ of the biased qubit-spin model obtained by the numerical HEOM method and the perturbative approach proposed in Ref.~\cite{28} for different parameters: $\lambda=5\gamma_{0}$ (numerical result: orange dotted line, perturbative theory: purple rectangles), $\lambda=2\gamma_{0}$ (numerical result: magenta dotdashed line, perturbative theory: blue diamonds) and $\lambda=0.5\gamma_{0}$ (numerical result: brown dashed line, perturbative theory: green five-point stars), other parameters are chosen as $\epsilon=0.85$, $\Delta=-0.3\epsilon$, $\gamma_{0}=0.02$, $\omega_{0}=0$ and the initial state is $|\psi(0)\rangle=|e\rangle$. (b) The same with (a) but $\gamma_{0}=0.05$. (c) The same with (a) but $\gamma_{0}=0.10$. (d) $\ln[\Gamma(\tau)]$ obtained by the numerical HEOM method versus $\tau$: $\lambda=10\gamma_{0}$ (blue solid line) and $\lambda=0.1\gamma_{0}$ (red dashed line) with $\gamma_{0}=1$, other parameters are the same with Fig.~\ref{fig:fig2}(a).}
\end{figure}

Next, we adopt the numerical HEOM method to compute the effective decay rate $\Gamma(\tau)$ in the case $\Delta\neq 0$ and compare our numerical result with that of the perturbative approach in Ref.~\cite{28}, where only the first and second order terms of the time-evolution operator are considered. This perturbative approach proposed in Ref.~\cite{28} allows us to obtain a general expression of the effective decay rate but only works in the weak-coupling regime.

In Figs.~\ref{fig:fig2} (a), (b) and (c), we display the effective decay rate $\Gamma(\tau)$ obtained by the numerical HEOM method and the perturbative approach for different system-bath coupling strengths, respectively. It is clear to see that these two approaches are in good agreement when system-bath coupling strength is weak. For the strong-coupling regime, though there is a small deviation between the two kinds of methods, the results calculated by the perturbative approach can still qualitatively follow these of the numerical HEOM method. As you can see from Figs.~\ref{fig:fig2} (a), (b) and (c), $\Gamma(\tau)$ now in general has multiple peaks which indicates that there are multiple Zeno-anti-Zeno transition phenomena occur in this biased qubit-boson model. The existence of multiple peaks can be significant for experiments, because now a (local) Zeno-anti-Zeno transition may be also observed using a relatively large measurement interval.

Now, we try to explore the relationship between the non-Markovianity and the QZE in quantum dissipative systems. From Fig.~\ref{fig:fig1}, one can find that the value of the effective decay rate's derivative at $\tau=0$, namely $\frac{d}{d\tau}\Gamma(\tau)|_{\tau\rightarrow 0}$, increases with the decrease of $\gamma_{0}/\lambda$ (throughout this paper, we fix the value of $\gamma_{0}$ and change the value of $\lambda$). In fact, in the short time regime, $\cos(\omega\tau)\simeq 1-\frac{1}{2}\omega^{2}\tau^{2}$, one can obtain the analytical expression of $\Gamma(\tau)$ in Eq.~\ref{Eq:Eq10}. Under this approximation, the quantum Zeno time is given by $\tau_{\mathrm{Z}}\simeq(\gamma_{0}\lambda)^{-1/2}$, where we have set $\omega_{0}=0$ for the sake of simplicity. This result indicates that the quantum Zeno time $\tau_{\mathrm{Z}}$ becomes short by decreasing the value of $\gamma_{0}/\lambda$. As our previous analysis, if the value of $\gamma_{0}/\lambda$ is very small, the decoherence dynamics is Markovian. Thus, we conclude that the quantum Zeno time $\tau_{\mathrm{Z}}$ in a Markovian bath is shorter than that of the non-Markovian bath.

The same result is also found in the biased qubit-boson model, due to the fact that $\Gamma(\tau)$ is very small when $\tau\rightarrow 0$, we plot the $\ln[\Gamma(\tau)]$ versus $\tau$ in Figs.~\ref{fig:fig2} (d). It is shown that the value of $\frac{d}{d\tau}\ln[\Gamma(\tau)]|_{\tau\rightarrow 0}$ in a Markovian bath is larger than that of the non-Markovian bath. Considering the fact that $\Gamma(\tau\rightarrow 0)$ in the Markovian bath is larger than that of the non-Markovian bath, one can demonstrate that the quantum Zeno time $\tau_{\mathrm{Z}}$ in a Markovian bath is shorter than that of the non-Markovian case in this biased qubit-boson model as well. In this sense, the non-Markovianity may prolong the quantum Zeno time. Our results suggest that the non-Markovianity may be favorable for the accessibility of the QZE in quantum dissipative systems.

\subsection{The biased qutrit-boson model}\label{sec:sec3b}

\begin{figure}
\centering
\includegraphics[angle=0,width=6cm]{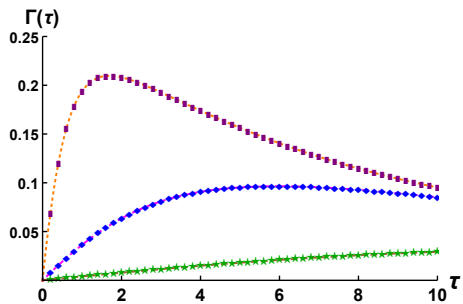}
\caption{\label{fig:fig3} Effective decay rate $\Gamma(\tau)$ of the pure dephasing qutrit-boson model obtained by the numerical HEOM method and the exact analytical expression for initial state $|\psi(0)\rangle=|\varsigma,J\rangle$ with different parameters: $\lambda=10\gamma_{0}$ (numerical result: orange dotted line, analytical result: purple rectangles), $\lambda=\gamma_{0}$ (numerical result: magenta dotdashed line, analytical result: blue diamonds) and $\lambda=0.1\gamma_{0}$ (numerical result: brown dashed line, analytical result: green five-point stars). Other parameters are chosen as $\epsilon=1$, $\gamma_{0}=0.2$, $\Delta=0$, $\omega_{0}=0$, $\phi_{0}=0$ and $\theta=\pi/2$.}
\end{figure}

In this subsection, we extend our study to a more general quantum dissipative system, in which a biased qutrit is coupled to a bosonic bath. The Hamiltonian of the quantum subsystem $\hat{H}_{s}$ and the operator coupled to the bath $f(\hat{s})$ are given by $\hat{H}_{s}=\epsilon \hat{J}_{z}+\Delta \hat{J}_{x}$ and $f(\hat{s})=2\hat{J}_{z}$, respectively. Operators $\hat{J}_{z}$ and $\hat{J}_{x}$ are defined by
\begin{equation*}
\hat{J}_{z}\equiv\left(
                   \begin{array}{ccc}
                     1 & 0 & 0 \\
                     0 & 0 & 0 \\
                     0 & 0 & -1 \\
                   \end{array}
                 \right);~~~\hat{J}_{x}\equiv\frac{1}{\sqrt{2}}\left(
                   \begin{array}{ccc}
                     0 & 1 & 0 \\
                     1 & 0 & 1 \\
                     0 & 1 & 0 \\
                   \end{array}
                 \right).
\end{equation*}
These matrices are expressed in the basis of $\{|J=1,m=1\rangle, |J=1,m=0\rangle, |J=1,m=-1\rangle\}$ where $|J,m\rangle$ are the eigenstates of $\hat{J}_{z}$ with $\hat{J}_{z}|J,m\rangle=m|J,m\rangle$. We can further generalize our analysis to the spin-$J$ ($J>1$) quantum dissipative systems which are also relevant to the two-component Bose-Einstein condensates~\cite{34}.

\begin{figure}
\centering
\includegraphics[angle=0,width=6cm]{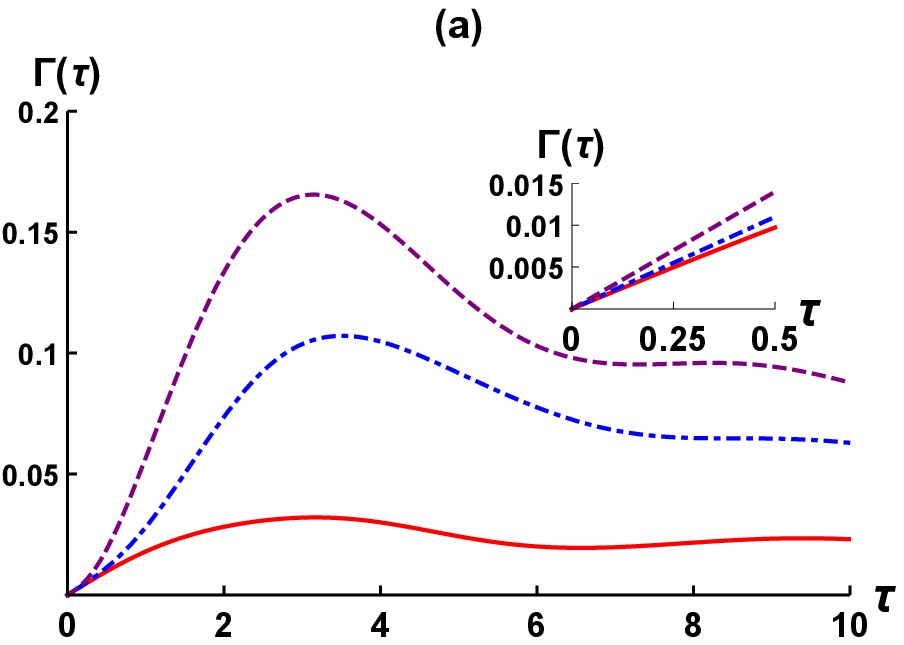}
\includegraphics[angle=0,width=6cm]{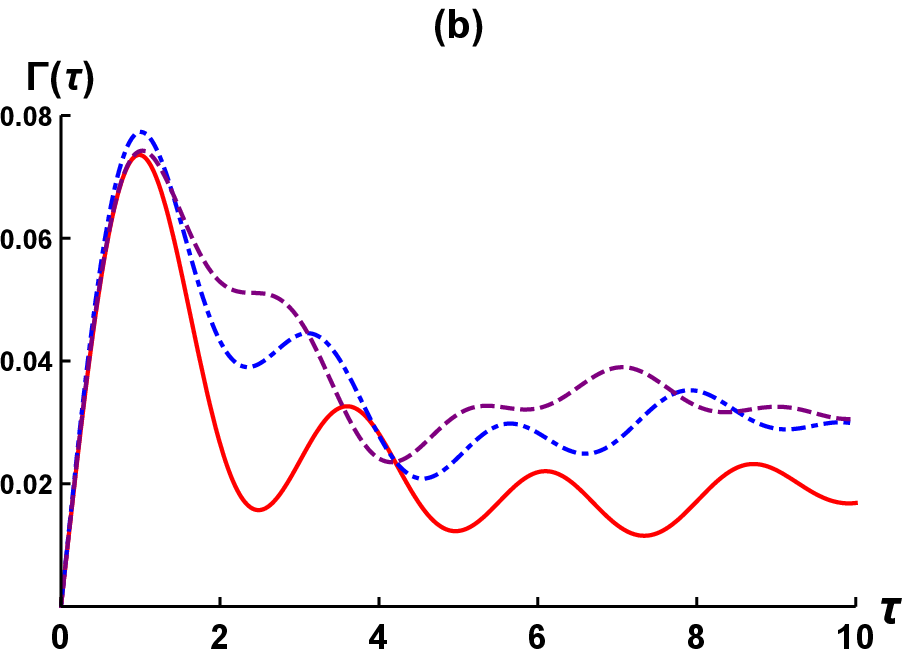}
\caption{\label{fig:fig4} (a) Effective decay rate $\Gamma(\tau)$ of the biased qutrit-spin model obtained by the numerical HEOM method with different spectral widths: $\lambda=10\gamma_{0}$ (purple dot-dashed line), $\lambda=\gamma_{0}$ (blue dashed line) and $\lambda=0.1\gamma_{0}$ (red solid line), other parameters are chosen as $\epsilon=1$, $\Delta=0.5\epsilon$, $\gamma_{0}=0.5$, $\omega_{0}=0$ and the initial state is $|\psi(0)\rangle=|J=1,m=1\rangle$. (b) The effective decay rate $\Gamma(\tau)$ obtained by the numerical HEOM method with different tunneling parameters: $\epsilon/\Delta=1.2$ (purple dot-dashed line), $\epsilon/\Delta=1.0$ (blue dashed line) and $\epsilon/\Delta=0.8$ (red solid line), other parameters are chosen as $\Delta=1$, $\lambda=5\gamma_{0}$, $\gamma_{0}=0.05$, $\omega_{0}=\sqrt{\epsilon^{2}+\Delta^{2}}$ and the initial state is $|\psi(0)\rangle=|J=1,m=1\rangle$.}
\end{figure}

For the case $\Delta=0$, this qutrit-boson model can be exactly solved. The reduced density matrix elements of the quantum subsystem are found to be~\cite{9}
\begin{equation}\label{Eq:Eq11}
\begin{split}
\varrho_{mm'}(t)=&\varrho_{mm'}(0)e^{-i\epsilon(m-m')t}e^{-i\phi(t)(m^{2}-m'^{2})}\\
&\times e^{-\kappa(t)(m-m')^{2}},
\end{split}
\end{equation}
where
\begin{equation*}
\phi(t)=4\int d\omega J_{\mathrm{L}}(\omega)\frac{\sin(\omega t)-\omega t}{\omega^{2}},
\end{equation*}
describes the phase diffusion effects induced by the bath in this model. In order to compare with the exact analytical results in Ref.~\cite{9}, we first take the initial state of the quantum subsystem as a standard SU(2) coherent state,
\begin{equation}\label{Eq:Eq12}
|\varsigma,J\rangle=(1+|\varsigma|^{2})^{-J}\sum_{m=-J}^{J}\sqrt{\left(
                                                                   \begin{array}{c}
                                                                     2J \\
                                                                     J+m \\
                                                                   \end{array}
                                                                 \right)
}\varsigma^{J+m}|J, m\rangle,
\end{equation}
where $\varsigma=e^{i\phi_{0}}\tan(\theta/2)$ with $\phi_{0}$ and $\theta$ being the parameters on the Bloch sphere. One can obtain the expression of the effective decay rate $\Gamma(\tau)$ of this initial state as follows~\cite{9}:
\begin{equation}\label{Eq:Eq13}
\begin{split}
\Gamma(\tau)=&-\frac{1}{\tau}\ln[(\frac{|\varsigma|}{1+|\varsigma|^{2}})^{4J}\sum_{m,m'}|\varsigma|^{2(m+m')}\left(
                                                                                                               \begin{array}{c}
                                                                                                                 2J \\
                                                                                                                 J+m \\
                                                                                                               \end{array}
                                                                                                             \right)
\\
&\times\left(
    \begin{array}{c}
      2J \\
      J+m' \\
    \end{array}
  \right)e^{-i\phi(\tau)(m^{2}-m'^{2})}e^{-\kappa(\tau)(m-m')^{2}}].
\end{split}
\end{equation}

In Fig.~\ref{fig:fig3}, we compare the exactly analytical results given by Eq.~\ref{Eq:Eq13} and the numerical results from the HEOM approach. A prefect agreement is found. We also find that $\Gamma(\tau)$ has single peak in our study which is different from that of Ref.~\cite{9}. Considering the fact that we chosen the Lorentz spectral function instead of the Ohmic spectrum in Ref.~\cite{9}, we conclude that the phenomenon of the occurrence of the multiple Zeno-anti-Zeno crossover phenomena may be sensitive to the characteristics of the bath density spectral function. Next, we consider a more general case, i.e., $\Delta\neq 0$; the value of the effective decay rate $\Gamma(\tau)$ is numerically computed by the HEOM method. We observe that $\Gamma(\tau)$ also has multiple peaks or multiple Zeno-anti-Zeno transitions at intermediate measurement intervals. As you can see from Fig.~\ref{fig:fig4}(b), this phenomenon can become more obvious by adjusting the ratio of $\epsilon/\Delta$ which indicates that the multiple-Zeno-anti-Zeno-transition phenomenon is sensitive to the value of the external bias field.

Similar to that of the qubit-boson model case, regardless of $\Delta=0$ or $\Delta\neq 0$, we find that the value of $\frac{d}{d\tau}\Gamma(\tau)|_{\tau\rightarrow 0}$ in the Markovian bath is larger than that of the non-Markovian bath. This result also suggests that the non-Markovian effects of the bath may prolong the quantum Zeno time and may add the possibility to realize the QZE in quantum dissipative systems.

\subsection{The finite temperature case}\label{sec:sec3c}

In this subsection, we investigate the relationship between the bath temperature and the QZE (QAZE) in quantum dissipative systems. We assume that the bath is now prepared in a thermal equilibrium state $\hat{\varrho}_{b}=\exp(-\beta\hat{H}_{b})/\mathrm{Tr}_{b}[\exp(-\beta\hat{H}_{b})]$ and the bath density spectral function $J(\omega)$ is the Ohmic spectrum with Drude cutoff
\begin{equation}\label{Eq:Eq14}
J_{\mathrm{O}}(\omega)=\frac{1}{\pi}\frac{2\chi\omega_{c}\omega}{\omega^{2}+\omega_{c}^{2}},
\end{equation}
where $\chi$ stands for the coupling strength between the quantum subsystem and its surrounding bath, parameter $\omega_{c}$ is the cutoff frequency. The reduced dynamics of the quantum subsystem can be numerically explored by the hierarchical equations given by Eq.~\ref{Eq:Eq20} in the Appendix. In Fig.~\ref{fig:fig5}, we display the effective decay rate $\Gamma(\tau)$ versus $\tau$ at different temperatures. As you can see from Fig.~\ref{fig:fig5}, the multiple Zeno-anti-Zeno crossover phenomena are observed in the finite temperature case as well. We also find that the value of $\frac{d}{d\tau}\Gamma(\tau)|_{\tau\rightarrow 0}$ becomes larger with the increase of the bath temperature regardless of the qubit-boson model or the qutrit-boson model. This result suggests that the high bath temperature may enhance the difficulty in realizing the QZE in quantum dissipative systems.

We would like to add some physical explanations about why this phenomenon occurs. When the bath temperature is very high, i.e. $\beta\rightarrow 0$, the bath correlation function $C_{\mathrm{O}}(t)$ of Eq.~\ref{Eq:Eq19} in the Appendix can be approximately replaced by the first term of the series as follows:
\begin{equation*}
C_{\mathrm{O}}(t)\simeq[\chi\omega_{c}\cot(\frac{\beta\omega_{c}}{2})-i\chi\omega_{c}]e^{-\omega_{c}t}\simeq(\frac{2\chi}{\beta}-i\chi\omega_{c})e^{-\omega_{c}t}.
\end{equation*}
In this approximation, the hierarchical equation given by Eq.~\ref{Eq:Eq20} in the Appendix is equivalent to the Markovian Zusman equation~\cite{35}. This result indicates that the non-Markovian characteristic becomes weak with the increase of the bath temperature. The same conclusion is also reported in Refs.~\cite{23,add1}. In Ref.~\cite{23}, the authors demonstrated that, by making use of the trace distance, the degree of non-Markovianity in the qubit-boson model almost equals zero when the bath temperature is very high. In Ref.~\cite{add1}, the authors found that the backflow of quantum information from the bath to the system is reduced, which means the decrease of non-Markovianity, when bath temperature increases. And according to our analysis in Subsec.~\ref{sec:sec3a}, the non-Markovianity may add the possibility to realize the QZE for a qubit or a qutrit coupled to a zero temperature Lorentzian environment. Therefore, for the qubit-boson and the qutrit-boson models considered in our paper, the quantum Zeno dynamics may become harder to realize with the increase of bath's temperature because a high temperature possibly reduces the non-Markovianity. In this sense, the result obtained in this subsection is consistent with our previous analysis in Subsec.~\ref{sec:sec3a}. Several previous studies have come to the same conclusion, for example, in Ref.~\cite{10}, the authors found that, for the pure dephasing qubit-boson model, the quantum Zeno time in terms of quantum Fisher information becomes smaller with the increase of the bath temperature, this result indicates that the increase of the bath temperature makes the quantum Zeno dynamics more difficult. However, considering the fact that our Hamiltonian of the quantum dissipative system is more general, this conclusion is more convincing.

\begin{figure}
\centering
\includegraphics[angle=0,width=6cm]{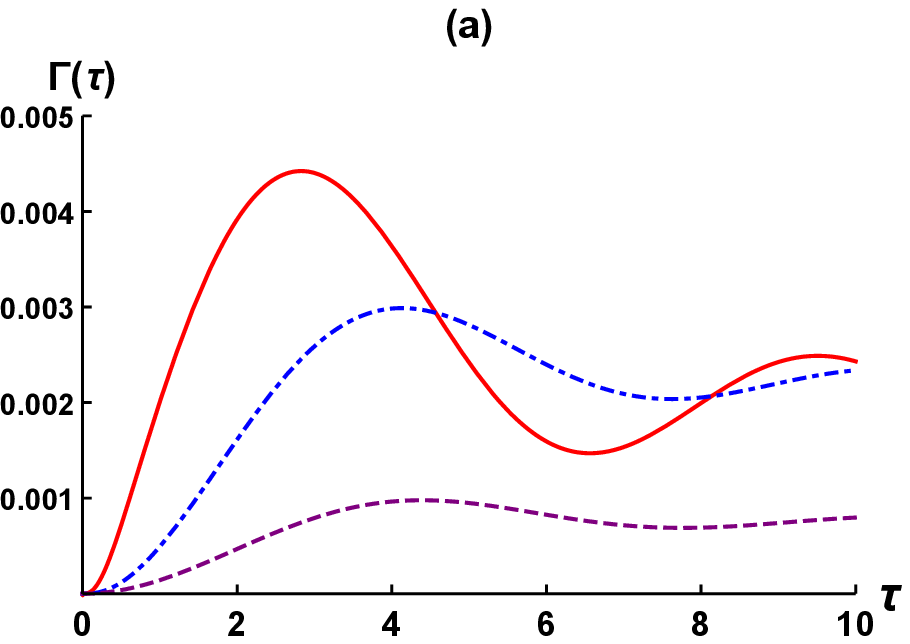}
\includegraphics[angle=0,width=6cm]{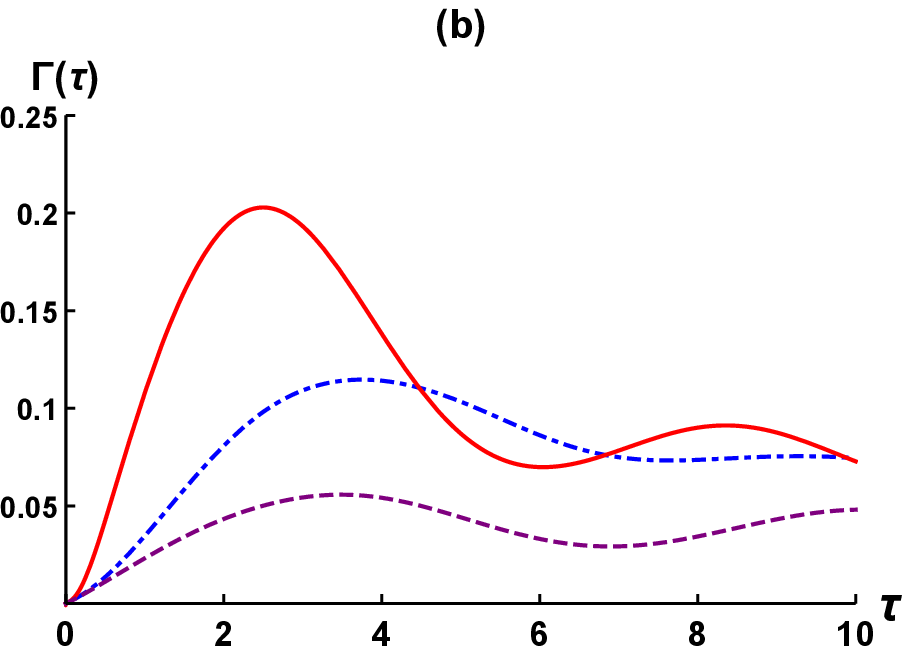}
\caption{\label{fig:fig5} (a) Effective decay rate $\Gamma(\tau)$ of the biased qubit-boson model obtained by the numerical HEOM method at different temperatures: $\beta\epsilon=0.5$ (purple dot-dashed line), $\beta\epsilon=0.1$ (blue dashed line) and $\beta\epsilon=0.01$ (red solid line), other parameters are chosen as $\epsilon=1$, $\Delta=-0.1\epsilon$, $\chi=0.05$, $\omega_{c}=10\epsilon$ and the initial state is $|\psi(0)\rangle=|e\rangle$. (b) The effective decay rate $\Gamma(\tau)$ of the biased qutrit-boson model obtained by the numerical HEOM method at different temperatures: $\beta\epsilon=1$ (purple dot-dashed line), $\beta\epsilon=0.15$ (blue dashed line) and $\beta\epsilon=0.01$ (red solid line), other parameters are chosen as $\epsilon=1$, $\Delta=0.5\epsilon$, $\chi=0.05$, $\omega_{c}=10\epsilon$ and the initial state is $|\psi(0)\rangle=|J=1,m=1\rangle$.}
\end{figure}

\section{Discussions and Conclusions}\label{sec:sec4}

Here, we would like to provide a possible physical explanation why the non-Markovianity may be favorable for the accessibility of the QZE in quantum dissipative systems. For the sake of simplicity, we restrict our discussion in the two-level quantum open system to the initial pure state. To establish the linkage between the non-Markovianity and the QZE in two-level quantum open systems, we first need to demonstrate the following \emph{lemma} which is very helpful for our analysis.

\emph{Lemma}. A larger value of  $D[\hat{\varrho}_{s}(t),\hat{\varrho}_{s}^{\bot}(0)]$ makes a longer quantum Zeno time $\tau_{\mathrm{Z}}$ in two-level quantum open systems, where $D[\hat{\rho},\hat{\varrho}]\equiv\frac{1}{2}\rm Tr[\sqrt{(\hat{\rho}-\hat{\varrho})^{\dagger}(\hat{\rho}-\hat{\varrho})}]$ denotes the trace distance~\cite{22} between quantum state $\hat{\rho}$ and $\hat{\varrho}$, $\hat{\varrho}^{\bot}$ refers to an arbitrary quantum state orthogonal to $\hat{\varrho}$.

\emph{Proof}. Assuming the initial state of a two-level quantum subsystem is given by $\hat{\varrho}_{s}(0)=|\varphi_{1}\rangle\langle\varphi_{1}|$, where $|\varphi_{1,2}\rangle$ are the basis vectors of a certain representation, then the reduced density matrix at time $t$ can be expressed in the following form: $\hat{\varrho}_{s}(t)=c_{11}(t)|\varphi_{1}\rangle\langle\varphi_{1}|+c_{12}(t)|\varphi_{1}\rangle\langle\varphi_{2}|+c_{12}^{*}(t)|\varphi_{2}\rangle\langle\varphi_{1}|+c_{22}(t)|\varphi_{2}\rangle\langle\varphi_{2}|$. The trace distance between $\hat{\varrho}_{s}(t)$ and $\hat{\varrho}_{s}^{\bot}(0)=|\varphi_{2}\rangle\langle\varphi_{2}|$ is given by $D[\hat{\varrho}_{s}(t),\hat{\varrho}_{s}^{\bot}(0)]=\sqrt{c_{11}^{2}(t)+|c_{12}(t)|^{2}}$. On the other hand, the survival probability of the initial state $\hat{\varrho}_{s}(0)$ can be expressed as $c_{11}(t)\simeq1-t^{2}/\tau_{\mathrm{Z}}^{2}$ in the short time region. Then, one can find that $D[\hat{\varrho}_{s}(t),\hat{\varrho}_{s}^{\bot}(0)]=\sqrt{(1-t^{2}/\tau_{\mathrm{Z}}^{2})^{2}+|c_{12}(t)|^{2}}$. From this expression, it is easy to see that, for a certain fixed time $t$, a larger value of $D[\hat{\varrho}_{s}(t),\hat{\varrho}_{s}^{\bot}(0)]$ induces a longer quantum Zeno time $\tau_{\mathrm{Z}}$ or makes the QZE easier to realize. In fact, considering the fact that the trace distance is a measure of distinguishability between two quantum states, this \emph{lemma} tells us a very intuitive physical result: a higher (or lower) distinguishability between $\hat{\varrho}_{s}(t)$ and $\hat{\varrho}_{s}^{\bot}(0)$ (or $\hat{\varrho}_{s}(0)$) makes the QZE easier to realize.

The physical meaning of the trace distance $D[\hat{\varrho}_{s}(t),\hat{\varrho}_{s}^{\bot}(0)]$ can be easily understood from the view of the exchange of quantum information~\cite{22,23,24,36}. We can define the change rate of the trace distance as $\varpi[\hat{\rho},\hat{\varrho}]\equiv dD[\hat{\rho},\hat{\varrho}]/dt$. When $\varpi[\hat{\varrho}_{s}(t),\hat{\varrho}_{s}^{\bot}(0)]<0$, $\hat{\varrho}_{s}(t)$ and $\hat{\varrho}_{s}^{\bot}(0)$ approach each other, and this can be understood as the quantum information flows from quantum subsystem to the bath; when $\varpi[\hat{\varrho}_{s}(t),\hat{\varrho}_{s}^{\bot}(0)]>0$, $\hat{\varrho}_{s}(t)$ and $\hat{\varrho}_{s}^{\bot}(0)$ are away from each other, and this can be interpreted as the quantum information flows back to the quantum subsystem, which is the typical character of the non-Markovianity~\cite{22,23,24,36}. Obviously, one can find that
\begin{equation}\label{Eq:Eq15}
\begin{split}
D[\hat{\varrho}_{s}(t),\hat{\varrho}_{s}^{\bot}(0)]=&D[\hat{\varrho}_{s}(0),\hat{\varrho}_{s}^{\bot}(0)]-\mathcal{I}_{\rm loss}[\hat{\varrho}_{s}(t),\hat{\varrho}_{s}^{\bot}(0)]\\
&+\mathcal{I}_{\rm gain}[\hat{\varrho}_{s}(t),\hat{\varrho}_{s}^{\bot}(0)],
\end{split}
\end{equation}
where $D[\hat{\varrho}_{s}(0),\hat{\varrho}_{s}^{\bot}(0)]=1$ due to the fact that $\hat{\varrho}_{s}(0)$ and $\hat{\varrho}_{s}^{\bot}(0)$ have orthogonal supports, $\mathcal{I}_{\rm loss}[\hat{\varrho}_{s}(t),\hat{\varrho}_{s}^{\bot}(0)]$ indicates the quantum information loss due to the dissipation and $\mathcal{I}_{\rm gain}[\hat{\varrho}_{s}(t),\hat{\varrho}_{s}^{\bot}(0)]$ refers to the quantum information gained from the bath during the time interval $[0,t]$. Their definitions are given by~\cite{36}
\begin{equation*}
\mathcal{I}_{\rm loss}[\hat{\varrho}_{s}(t),\hat{\varrho}_{s}^{\bot}(0)]\equiv-\int_{\varpi<0}dt\varpi[\hat{\varrho}_{s}(t),\hat{\varrho}_{s}^{\bot}(0)],
\end{equation*}
\begin{equation*}
\mathcal{I}_{\rm gain}[\hat{\varrho}_{s}(t),\hat{\varrho}_{s}^{\bot}(0)]\equiv\int_{\varpi>0}dt\varpi[\hat{\varrho}_{s}(t),\hat{\varrho}_{s}^{\bot}(0)].
\end{equation*}
In a Markovian process, the bath is memoryless and there is no feedback information flows from bath to the quantum subsystem, i.e. $\mathcal{I}_{\rm gain}[\hat{\varrho}_{s}(t),\hat{\varrho}_{s}^{\bot}(0)]=0$ for all the time interval $[0,t]$. In this case, the trace distance $D[\hat{\varrho}_{s}(t),\hat{\varrho}_{s}^{\bot}(0)]$ monotonically decreases, $\hat{\varrho}_{s}(t)$ and $\hat{\varrho}_{s}^{\bot}(0)$ become more and more ``similar". On the contrary, for a non-Markovian process, the feedback quantum information flows from bath to the quantum subsystem may increase the distinguishability between $\hat{\varrho}_{s}(t)$ and $\hat{\varrho}_{s}^{\bot}(0)$, namely the value of $D[\hat{\varrho}_{s}(t),\hat{\varrho}_{s}^{\bot}(0)]$ becomes large. In this sense, the trace distance $D[\hat{\varrho}_{s}(t),\hat{\varrho}_{s}^{\bot}(0)]$ in a non-Markovian bath may be larger than the trace distance $D[\hat{\varrho}_{s}(t),\hat{\varrho}_{s}^{\bot}(0)]$ in a Markovian bath. According to the \emph{lemma}, this non-monotonic behavior of quantum information flow may induce a longer quantum Zeno time, or in other words, the non-Markovianity may be favorable for the accessibility of the QZE in quantum open systems.

We want to emphasize that $\mathcal{I}_{\rm gain}[\hat{\rho},\hat{\varrho}]$ is not a rigorous physical quantity to characterize the non-Markovianity in quantum open systems. $\mathcal{I}_{\rm gain}[\hat{\varrho}_{s}(t),\hat{\varrho}_{s}^{\bot}(0)]$ only quantifies the quantum information, which flows from the bath back to the quantum subsystem, with respect to the standard state $\hat{\varrho}_{s}^{\bot}(0)$ during the time interval $[0,t]$. In this case, the value of $\mathcal{I}_{\rm gain}[\hat{\varrho}_{s}(t),\hat{\varrho}_{s}^{\bot}(0)]$ strongly lies on the chosen of the standard state $\hat{\varrho}_{s}^{\bot}(0)$ or the initial state $\hat{\varrho}_{s}(0)$. However, any rigorous measure of the non-Markovianity of quantum open system should focus on the decoherence mechanism itself and be independent of the initial state. Meanwhile, as shown in Ref.~\cite{36}, $\mathcal{I}_{\rm gain}[\hat{\rho},\hat{\varrho}]$ still can be regarded as a modified measure of non-Markovianity and may give the same results with the measure of non-Markovianity given in Ref.~\cite{22} for some physical models.

In our theoretical formalism, we assume that all the measurements are ideal and instantaneous. After each measurement, the measured quantum subsystem is completely collapsed to its initial state. This postulation of wave-packet collapse is comprehensively adopted in most of the previous studies~\cite{1,2,9,10,11,12,28,29} and very beneficial to our numerical approach. However, it is necessary to point out that the QZE and the QAZE are independent of the postulation of wave-packet collapse. In Ref.~\cite{37}, the authors provide a description of the QZE and the QAZE in quantum dissipative systems without using the wave-packet collapse postulation and discuss the effect of the non-demolition measurement on the QZE and the QAZE. Furthermore, many literatures have shown that the effect of measurement can be replaced by a continuous strong coupling between the to-be-measured system and an auxiliary apparatus~\cite{38}. Nevertheless, neither of these two formalisms is suitable for a numerical study of the QZE and the QAZE in quantum dissipative systems, it would be very interesting to numerically explore the QZE and the QAZE in quantum dissipative systems beyond the wave-packet collapse postulation.

The QZE (QAZE) has become a focus of attention not only because of its foundational implications about the quantum mechanics, but also because it may be exploited to explore some potential applications in the quantum control~\cite{39}, the communication complexity problem~\cite{40}, and the quantum state protection or preparation~\cite{41}. To achieve any of the above potential applications, the noisy bath can not be ignored. This is the main reason why we consider the effects of the non-Markovian noise and the bath temperature on the QZE (and QAZE). Our work is a step forward in a physical insight into understanding of the QZE and the QAZE in quantum dissipative systems.

In conclusion, we adopt a numerical HEOM method to investigate the QZE and the QAZE in quantum dissipative systems. This numerical approach allows us to study the dynamical behavior of a quantum dissipative system without the usual Markovian approximation, the rotating-wave approximation, and the perturbative approximation. We investigate the effective decay rates of a biased qubit-boson model and a biased qutrit-boson model at both zero and finite temperatures. The multiple Zeno-anti-Zeno crossover phenomena are found in these quantum dissipative systems. It is necessary to point out that the existence of multiple peaks can be significant for experiments, because it allows us to observe the Zeno-anti-Zeno transition by a relatively large measurement interval. We also find that the value of the quantum Zeno time in a Markovian regime is shorter than that of the non-Markovian regime. This result suggests that the non-Markovianity of the bath may be favorable for the accessibility of the QZE in quantum dissipative systems. Moreover, for the finite-temperature case, it is shown that the value of the quantum Zeno time at high bath temperature is shorter than that of the low temperature case. This result implies that the high bath temperature may add the difficulty in observing the QZE in quantum dissipative systems. Finally, due to the generality of the spin-boson model, we expect our results to be of interest for a wide range of experimental applications in quantum computation and quantum information processing.

\section{Acknowledgments}\label{sec:secack}

W. W. wishes to thank Dr. D.-W. Luo, Prof. H. Zheng, Prof. X.-D. Hu and Prof. J.-Q. You for many useful discussions. W. W. and H.-Q. L. acknowledge the support from NSAF U1530401 and the computational resource from the Beijing Computational Science Research Center.

\section{Appendix: Methodology}\label{sec:secapp}

In this appendix, we briefly outline how to use the HEOM method to study the reduced dynamics of a biased qubit coupled to a bosonic bath. To demonstrate the validity of the numerical results obtained by the HEOM method, we make some careful comparisons between our numerical results and the analytical results from the generalized Silbey-Harris transformation~\cite{12,13,14}.

The bath correlation function $C(t)$ of a general spin-boson model is given by
\begin{equation*}
\begin{split}
C(t)\equiv& \mathrm{Tr}_{b}[\hat{g}_{b}(t)\hat{g}_{b}(0)\hat{\varrho}_{b}]\\
=&\int d\omega J(\omega)[\coth(\frac{\beta\omega}{2})\cos(\omega t)-i\sin(\omega t)],
\end{split}
\end{equation*}
where $\hat{g}_{b}(t)\equiv\sum_{k}g_{k}(a_{k}^{\dagger}e^{i\omega_{k}t}+a_{k}e^{-i\omega_{k}t})$. If the bath correlation function can be expressed as a sum of exponential functions, i.e.,
\begin{equation}\label{Eq:Eq16}
C(t)=\sum_{j}\zeta_{j} e^{-\upsilon_{j} t},
\end{equation}
where $\zeta_{j}$ and $\upsilon_{j}$ are assumed to be complex numbers for the generality, the hierarchical equations can be derived by making use of the Feynman-Vernon influence functional approach~\cite{17,18} or the superoperator technique~\cite{19,20,42}. Equation~\ref{Eq:Eq16} is the key condition to perform the HEOM method. For the Lorentz spectrum, the bath correlation function at zero temperature is given by
\begin{equation}\label{Eq:Eq17}
C_{\mathrm{L}}(t)=\frac{1}{2}\gamma_{0}\lambda e^{-(\lambda+i\omega_{0}) t}.
\end{equation}
This is the simplest case of Eq.~\ref{Eq:Eq16}. Following procedures shown in Ref.~\cite{20}, one can obtain the hierarchy equations of the reduced quantum subsystem as follows:
\begin{equation}\label{Eq:Eq18}
\begin{split}
\frac{d}{dt}\hat{\rho}_{\vec{l}}(t)=&(-i\hat{H}_{s}^{\times}-\vec{l}\cdot\vec{\nu})\hat{\rho}_{\vec{l}}(t)+\hat{\Phi}\sum_{p=1}^{2}\hat{\rho}_{\vec{l}+\vec{e}_{p}}(t)\\
&+\sum_{p=1}^{2}l_{p}\hat{\Psi}_{p}\hat{\rho}_{\vec{l}-\vec{e}_{p}}(t),
\end{split}
\end{equation}
where $\vec{l}=(l_{1},l_{2})$ is a two-dimensional index, $\vec{e}_{1}=(1,0)$, $\vec{e}_{2}=(0,1)$, and $\vec{\nu}=(\lambda-i\omega_{0},\lambda+i\omega_{0})$ are two-dimensional vectors, two superoperators $\hat{\Phi}$ and $\hat{\Psi}_{p}$ are defined as follows:
\begin{equation*}
\hat{\Phi}=-if(\hat{s})^{\times},~~~\hat{\Psi}_{p}=\frac{i}{4}\gamma_{0}\lambda[(-1)^{p}f(\hat{s})^{\circ}-f(\hat{s})^{\times}],
\end{equation*}
with $\hat{X}^{\times}\hat{Y}\equiv[\hat{X},\hat{Y}]=\hat{X}\hat{Y}-\hat{Y}\hat{X}$ and $\hat{X}^{\circ}\hat{Y}\equiv\{\hat{X},\hat{Y}\}= \hat{X}\hat{Y}+\hat{Y}\hat{X}$. The same hierarchical equations can be also derived by making use of the stochastic decoupling scheme proposed by Shao et al.~\cite{27}.

The initial-state conditions of the auxiliary matrices are $\hat{\rho}_{\vec{l}=\vec{0}}(0)=\hat{\rho}_{s}(0)$ and $\hat{\rho}_{\vec{l}\neq \vec{0}}(0)=0$, where $\vec{0}=(0,0)$ is a two-dimensional zero vector. For numerical simulations, we need to truncate the number of hierarchical equations for a sufficiently large integer $L$, which means all the terms of $\hat{\rho}_{\vec{l}}(t)$ with $l_{1}+l_{2}>L$ are set to be zero. Then terms of $\hat{\rho}_{\vec{l}}(t)$ with $l_{1}+l_{2}\leq L$ form a closed set of differential equations.

In the finite-temperature case, the bath correlation function $C(t)$ does not satisfy the condition to perform the HEOM method for the Lorentz spectrum. Thus we assume that bath density spectral function is the Ohmic spectrum with Drude cutoff, namely $J(\omega)=J_{\mathrm{O}}(\omega)$, for the finite temperature case. The bath correlation function is then given by~\cite{17,18,19}
\begin{equation}\label{Eq:Eq19}
C_{\mathrm{O}}(t)=\sum_{j=0}^{\infty}\zeta_{j}e^{-\upsilon_{j}t},
\end{equation}
where $\upsilon_{j}=\omega_{c}\delta_{0j}+2j\pi(1-\delta_{0j})/\beta$ denotes the $j$th Matsubara frequency and
\begin{equation*}
\zeta_{j}=\frac{4\chi\omega_{c}}{\beta}\frac{\upsilon_{j}}{\upsilon_{j}^{2}-\omega_{c}^{2}}(1-\delta_{0j})+[\chi\omega_{c}\cot(\frac{\beta\omega_{c}}{2})-i\chi\omega_{c}]\delta_{0j},
\end{equation*}
are the expansion coefficients. With the help of the bath correlation function given by Eq.~\ref{Eq:Eq19}, the hierarchical equations at finite temperature can be obtained as follows~\cite{17,18,19}:
\begin{equation}\label{Eq:Eq20}
\begin{split}
\frac{d}{dt}\hat{\rho}_{\vec{\ell}}(t)=&(-i\hat{H}_{s}^{\times}-\vec{\ell}\cdot\vec{\mu})\hat{\rho}_{\vec{\ell}}(t)\\
&-(\frac{2\chi}{\beta\omega_{c}}-i\chi-\sum_{q=0}^{\varepsilon}\frac{\zeta_{q}}{\upsilon_{q}})f(\hat{s})^{\times}f(\hat{s})^{\times}\hat{\rho}_{\vec{\ell}}(t)\\
&+\hat{\Phi}\sum_{q=0}^{\varepsilon}\hat{\rho}_{\vec{\ell}+\vec{e}_{q}}(t)+\sum_{q=0}^{\varepsilon}\ell_{q}\hat{\Theta}_{q}\hat{\rho}_{\vec{l}-\vec{e}_{q}}(t),
\end{split}
\end{equation}
where $\vec{\ell}=(\ell_{0},\ell_{1},\ell_{2},...,\ell_{\varepsilon})$ is a $(\varepsilon+1)$-dimensional index, $\vec{e}_{q}=(0,0,0,...1_{q},...,0)$ and $\vec{\mu}=(\upsilon_{0},\upsilon_{1},\upsilon_{2},...,\upsilon_{\varepsilon})$ are $(\varepsilon+1)$-dimensional vectors and $\varepsilon$ is the cutoff number of the Matsubara frequency. The superoperator $\hat{\Theta}_{q}$ is defined as
\begin{equation*}
\hat{\Theta}_{q}=-i[\zeta_{q}^{R}f(\hat{s})^{\times}+i\zeta_{q}^{I}f(\hat{s})^{\circ}],
\end{equation*}
where $\zeta_{q}^{R}$ and $\zeta_{q}^{I}$ are the real and imaginary parts of $\zeta_{q}$.

If the bath temperature is very high, i.e. $\beta\rightarrow 0$, or the value of $\varepsilon$ is very large, i.e. $\varepsilon\rightarrow\infty$, the term of
\begin{equation*}
(\frac{2\chi}{\beta\omega_{c}}-i\chi-\sum_{q=0}^{\varepsilon}\frac{\zeta_{q}}{\upsilon_{q}})f(\hat{s})^{\times}f(\hat{s})^{\times}\hat{\rho}_{\vec{\ell}}(t),
\end{equation*}
can be approximatively neglected~\cite{18} and Eq.~\ref{Eq:Eq20} reduces to the hierarchical equations in Ref.~\cite{27}. This approximation is reliable when the bath temperature is not very low. The initial-state conditions of the auxiliary matrices are $\hat{\rho}_{\vec{\ell}=\vec{\mathbf{0}}}(0)=\hat{\rho}_{s}(0)$ and $\hat{\rho}_{\vec{\ell}\neq \vec{\mathbf{0}}}(0)=0$, where $\vec{\mathbf{0}}=(0,0,0...,0)$ is a $(\varepsilon+1)$-dimensional zero vector. In this paper, we keep on adding the number of the differential equations until the final result converges. It is necessary to point out that the HEOM approach is independent of the usual Markovian approximation, the rotating-wave approximation, and the perturbative approximation; in this sense, it can be regarded as a rigorous numerical method.

\begin{figure}
\centering
\includegraphics[angle=0,width=4cm]{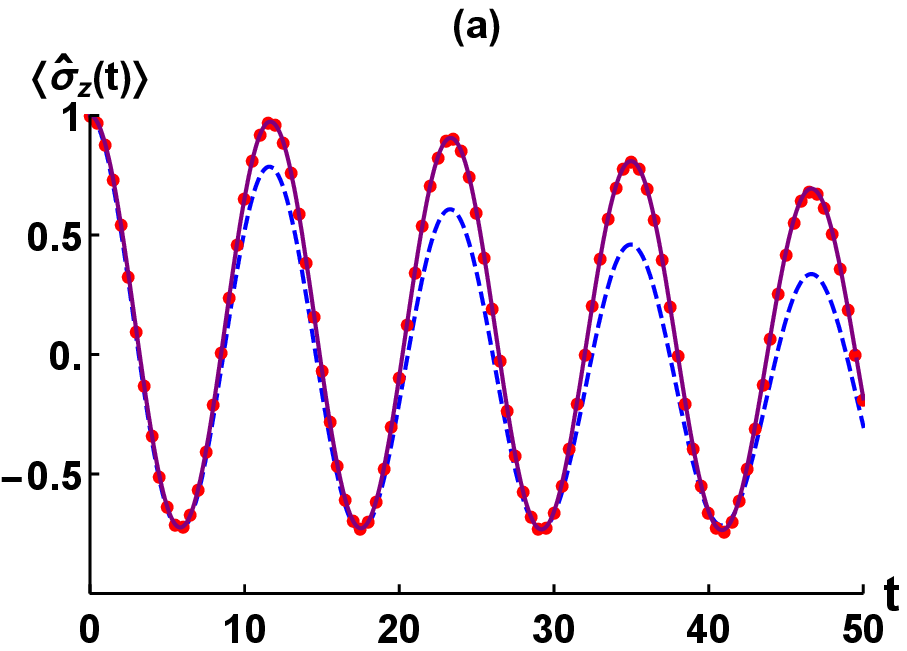}
\includegraphics[angle=0,width=4cm]{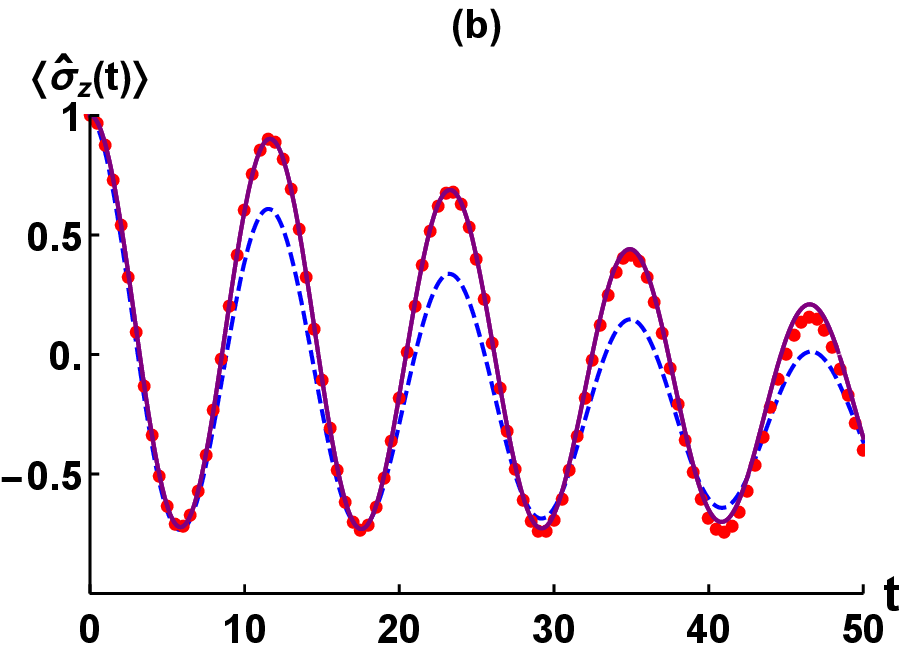}
\includegraphics[angle=0,width=4cm]{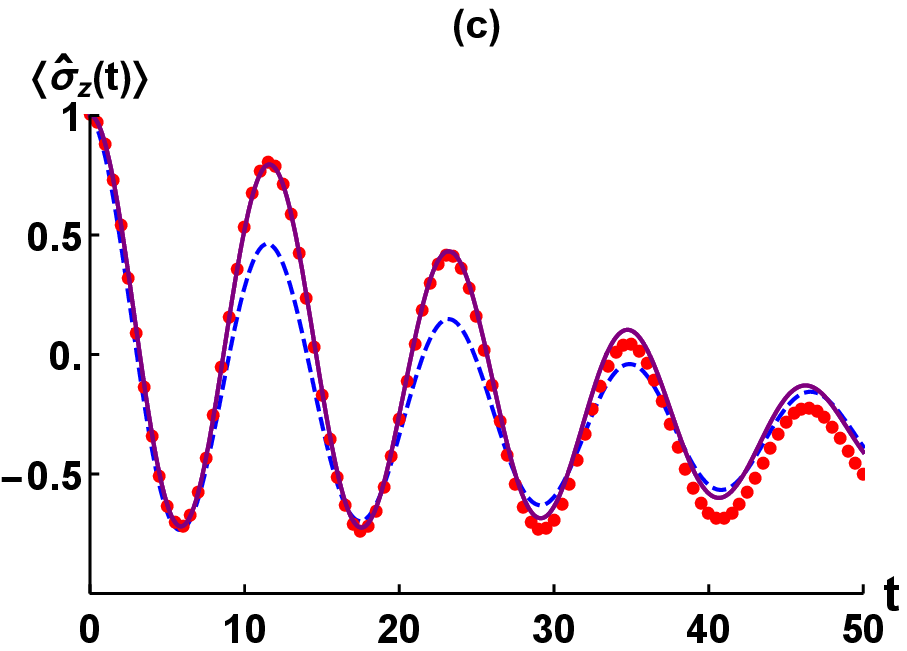}
\includegraphics[angle=0,width=4cm]{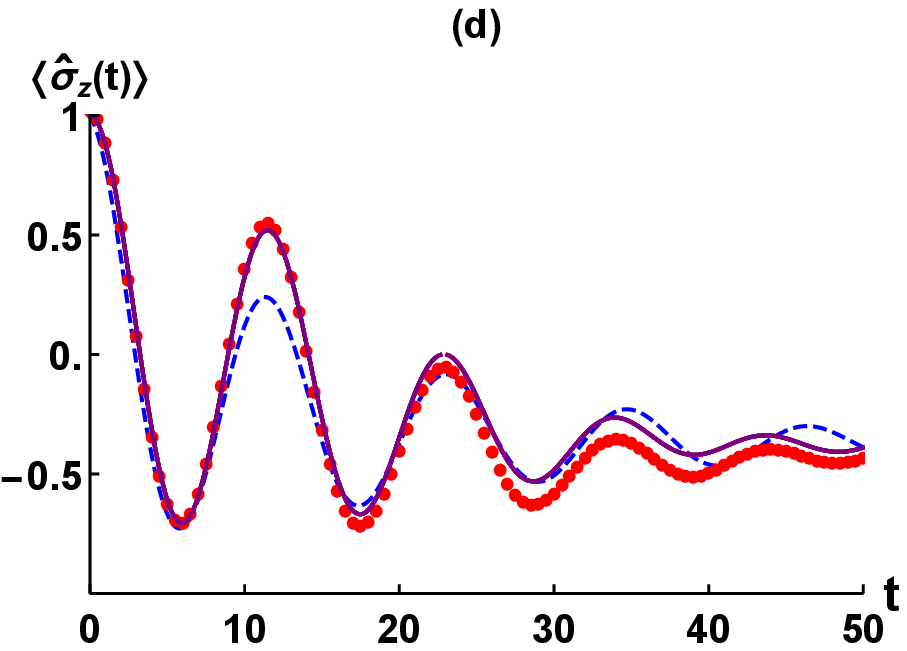}
\caption{\label{fig:fig6} Time evolution of the population difference $\langle\hat{\sigma}_{z}(t)\rangle$ for $\hat{H}_{s}=\frac{\epsilon}{2}\hat{\sigma}_{z}-\frac{\Delta}{2}\hat{\sigma}_{x}$ and $f(\hat{s})=\hat{\sigma}_{z}$ with different coupling parameters: (a) $\gamma_{0}=0.2\Delta$, (b) $\gamma_{0}=0.4\Delta$, (c) $\gamma_{0}=0.6\Delta$ and (d) $\gamma_{0}=1.0\Delta$. The purple solid lines are the numerical results obtained by the HEOM method, the red circles denote the results obtained by the generalized Silbey-Harris transformation approach and the blue dashed lines represent the Born-Markov results from Ref.~\cite{14}. Other parameters are chosen as $\epsilon=0.2$, $\Delta=2.5\epsilon$, $\lambda=0.2\gamma_{0}$ and $\omega_{0}=\sqrt{\epsilon^{2}+\Delta^{2}}$.}
\end{figure}

In order to verify the feasibility and the validity of the HEOM method, we make a comparison between the result obtained by the numerical HEOM method and that of the analytical generalized Silbey-Harris transformation approach~\cite{12,13,14}. In Fig.~\ref{fig:fig6}, we display the dynamics of the population difference $\langle\hat{\sigma}_{z}(t)\rangle\equiv \mathrm{Tr}[\hat{\varrho}_{s}(t)\hat{\sigma}_{z}]$ of the biased qubit-boson model, which is a very common quantity of interest in experiments. Here we assume that the initial state of the biased qubit is $\hat{\varrho}_{s}(0)=|e\rangle\langle e|$, and the bath is initially prepared in its Fock-vacuum state $\bigotimes_{k}|0_{k}\rangle$ with Lorentzian spectrum.

In Fig.~\ref{fig:fig6}(a), we consider the value of the system-bath coupling strength as $\gamma_{0}=0.2\Delta$, which is weak enough for this model, and compare the numerical results obtained by the HEOM approach and analytical results from the generalized Silbey-Harris transformation method. Good agreement is found between results from the two different approaches. For the strong-coupling regime, such as $\gamma_{0}=\Delta$ in Fig.~\ref{fig:fig6}(d), a deviation is found between the results calculated by the HEOM approach and the generalized Silbey-Harris transformation method. However, the HEOM results are believed to be more reliable, because the generalized Silbey-Harris transformation method neglects the higher-order terms of the system-bath coupling strength which is invalid in strong-coupling regime. In the moderately strong-coupling regime, such as $\gamma_{0}=0.4\Delta$ in Fig.~\ref{fig:fig6}(b) and $\gamma_{0}=0.6\Delta$ in Fig.~\ref{fig:fig6}(c), one can find that the results of the generalized Silbey-Harris transformation method are still in qualitative agreement with those of the numerical HEOM method. Moreover, we also compare these results with that of the Born-Markov approximation and find that the Born-Markov approximation gives a relatively large deviation in the population dynamics regardless of the weak-coupling or the strong-coupling regimes. In this sense, the generalized Silbey-Harris transformation method captures the dynamics of the biased qubit-boson model when the system-bath coupling is not too strong. On the contrary, the usual Born-Markov theory may give a qualitatively incorrect conclusion as a result of neglecting the feedback action of the bath with memory.

\end{document}